# Influence of the Metals and Ligands in Dinuclear Complexes on Phosphopeptide Sequencing by Electron Transfer Dissociation Tandem Mass Spectrometry

Daiki Asakawa[1*], Akio Miyazato[2], Frédéric Rosu[3] and Valérie Gabelica[4]

1. National Institute of Advanced Industrial Science and Technology (AIST), Tsukuba Central 2, 1-1-1 Umezono, Tsukuba, Ibaraki, Japan

2. Center for Nano Materials and Technology, Japan Advanced Institute of Science and Technology, 1-1 Asahidai, Nomi, Ishikawa, Japan

3. CNRS, INSERM, Univ. Bordeaux, Institut Européen de Chimie et Biologie (IECB, UMS3033, US001), 2 rue Robert Escarpit, 33607 Pessac, France.

4. Univ. Bordeaux, INSERM, CNRS, Laboratoire Acides Nucléiques Régulations Naturelle et Artificielle (ARNA, U1212, UMR5320), IECB, 2 rue Robert Escarpit, 33607 Pessac, France.

Corresponding to: Daiki Asakawa, Email: d.asakawa@aist.go.jp, TEL: +81-029-861-0586

National Institute of Advanced Industrial Science and Technology (AIST), Tsukuba Central 2, Umezono 1-1-1, Tsukuba, Ibaraki, 305-8568, JAPAN

ORCID: Daiki Asakawa: 0000-0002-9357-8420

Frédéric Rosu: 0000-0003-3674-7539

Valérie Gabelica: 0000-0001-9496-0165

# ABSTRACT

Phosphorylation is one of the most important protein modification, and electron transfer dissociation tandem mass spectrometry (ETD-MS/MS) is a potentially useful method for sequencing of phosphopeptide, including determination of phosphorylation site. Notably, ETD-MS/MS typically provide useful information when precursor contains more than three positive charges and has not yet closed as the method for large-scale phosphopeptide analysis, due to the difficulty of the production for acidic phopeptides having more than three positive charges. To increase the charge state of phosphopeptides, we have used the dinuclear metal complexes, which selectively bound to phosphate group in phosphopeptide with addition of positive charge(s). The dinuclear copper, zinc and gallium complexes have been tested and the type of metal present in the complex strongly affected to the affinity of phosphorylated compound and their ETD fragmentation. The dinulcear copper complex interact weakly with phosphate group and ETD induced peptide fragmentation is largely suppressed by the presence of $Cu^{2+}$, which worked as an electron trap. The dinulcear gallium complex strongly bound to phosphate group. However, the ligand binding to gallium acted as an electron trap and the presence of dinulcear gallium complex in precursor for ETD-MS/MS hampers the sequencing of phosphopeptide, as in the case of dinulcear copper complexes. In contrast, dinulcear zinc complexes efficiently bind to phosphopeptides with increase in the charge state, facilitating phosphopeptide sequencing by ETD-MS/MS. The fragmentation of ligand and peptide backbone in dinulcear zinc-phosphopeptide complex are competitively induced by ETD. These processes are influenced by ligand structure and the detailed ETD fragmentation pathway were investigated using density functional theory calculations.

Keywords: Radical site, Phosphorylation site, Density functional theory, Reduction

# INTRODUCTION

The reversible phosphorylation of proteins is a common covalent protein modification,[1-2] and electrospray ionization (ESI)-based tandem mass spectrometry (MS/MS) is currently the dominant method for the analysis of protein phosphorylation. In particular, collision-induced dissociation (CID) has been widely used to analyze the peptide sequence by MS/MS. In the CID of phosphorylated peptides, the loss of the phosphate groups (80 and 98 Da) can be used as a specific marker for phosphorylated peptide identification.[3-4] Therefore, the screening of phosphorylated peptides can be achieved by CID-MS/MS. However, the determination of the phosphorylated sites in peptides is hindered by the substantial loss of the phosphate groups because most of the fragment ions are observed in a non-phosphorylated form in CID-MS/MS spectra.[3, 5-6]

Fragmentation methods involving the electron association of multiply charged peptides, such as electron capture dissociation (ECD)[7] and electron transfer dissociation (ETD)[8], have been used as alternative methods to CID.[9-11] Because the ECD/ETD of phosphorylated peptides results in fragment ions arising from the fragmentation of the peptide backbone without losing phosphoric acid, the determination of the phosphorylation site is possible. Regarding the ECD/ETD mechanism, electron attachment/transfer occurs competitively at positively charged sites[12] and at the π* antibonding orbital of peptide bond[13-15] in multiply charged peptides, leading to N−Cα bond cleavage through aminoketyl radical intermediates. Although one of the most promising applications of ECD/ETD is phosphopeptide sequencing, it has not yet become the method of choice for large-scale phosphopeptide analysis because of the low fragmentation efficiency for tryptic phosphopeptides.[16]

Employing ions with a higher charge state as the precursors for ECD/ETD is a way to dramatically improve the fragmentation efficiency by increasing the yield of reactive radical species, and thereby improve the sequence coverage.[17-19] Here, we investigated how to increase the charge state of precursor ions to improve the quality of ETD mass spectra. In particular, the complexation of peptides with metal ions has been demonstrated to increase the charge state of peptides and various metal cations have been used as the cationization adductive for ETD-MS/MS experiments.[20-24] With regards to the ETD processes, the metal−peptide complex undergoes either N–Cα bond cleavage or metal cation reduction, according to the electrochemical properties of metal cations in the precursor. The metal cation having closed shell, such as monovalent alkali metal cations, divalent alkali earth and group XII metal cations, and trivalent group XIII metal cations are barely reduced by ETD-MS/MS condition.[25-28] In consequence, the use of metal cation having closed shell promotes the N–Cα bond cleavage and provides more informative MS/MS mass spectra compared to those obtained from protonated precursors with lower charge states.[21-22, 29] In contrast, transition metal cations with a partially filled d orbital shell, such as $Co^{2+}$, $Ni^{2+}$, and $Cu^{2+}$, are reduced by ETD and the fragmentation is mainly induced by the substantial excitation attributable to the reduction energy of the metal cation in the complex.[30] Although $Cd^{2+}$ and $Hg^{2+}$ are closed shelled metal cation, $Cd^{2+}$- and $Hg^{2+}$-peptide complexes undergoes metal cation reduction by ETD, probably due to complex electron energy level of $Cd^{2+}$ and $Hg^{2+}$.[22, 31] The results suggest that the period 5, 6 and 7 metal cations were susceptible to undergo the reduction by ETD. Therefore, alkali metals, alkaline earth metals, $Al^{3+}$, $Zn^{2+}$ and $Ga^{3+}$ are suggested to be suitable adductive for peptide sequencing by ETD-MS/MS. However, the yield of metal-peptide complex is usually low when the free metal cation was used.[21-22, 30]

To increase the yield of metal-phosphopeptide complex, we have recently used a dinuclear zinc complex, $(Zn_2L1)^{3+}$ (L1 = alkoxide form of 1,3-bis[bis(pyridin-2-ylmethyl)amino]propan-2-ol), also known as the phosphate capture molecule "phos-tag",[32-34] as an ionizing reagent.[16] The molecular structure of $(Zn_2L1)^{3+}$ is shown in Scheme 1a. Because $(Zn_2L1)^{3+}$ selectively binds with phosphorylated peptides with the addition of a positive charge per phosphate group, $(Zn_2L1)^{3+}$-aided ESI improves the ionization yield of phosphopeptides present in the protein digest,[16] and allowed us to sequence tryptic phosphopeptide by ETD-MS/MS. However, $(Zn_2L1)^{3+}$-aided ETD-MS/MS had one disadvantage: the fragmentation of $(Zn_2L1)^{3+}$ was also induced by ETD and competed with phosphopeptide backbone fragmentation. Therefore, the development of a new metal complex to replace $(Zn_2L1)^{3+}$ is necessary to suppress the ligand dissociation and to preferentially form the fragments arising from N–Cα bond cleavage.

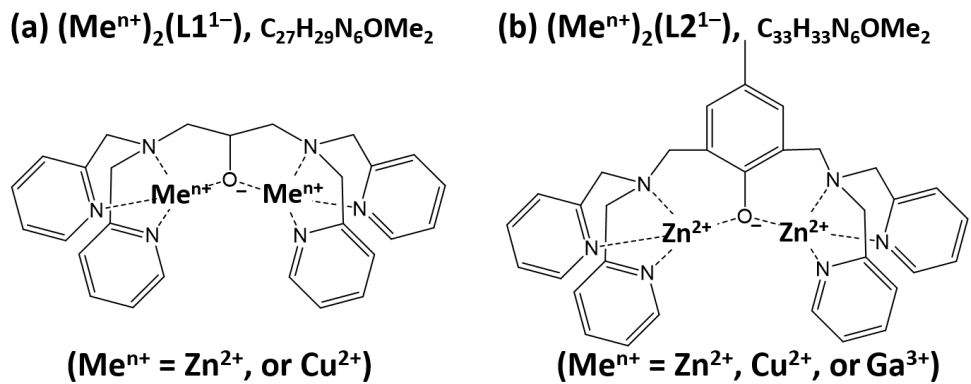

**(a) (Me$^{n+}$)$_2$(L1$^{1-}$), C$_{27}$H$_{29}$N$_6$OMe$_2$**

(Me$^{n+}$ = Zn$^{2+}$, or Cu$^{2+}$)

**(b) (Me$^{n+}$)$_2$(L2$^{1-}$), C$_{33}$H$_{33}$N$_6$OMe$_2$**

(Me$^{n+}$ = Zn$^{2+}$, Cu$^{2+}$, or Ga$^{3+}$)

**Scheme 1. Structure of dinuclear complexes (a) (Me$^{n+}$)$_2$(L1$^{-1}$) and (b) (Me$^{n+}$)$_2$(L2$^{-1}$), where Me$^{n+}$ is Zn$^{2+}$, Cu$^{2+}$, or Ga$^{3+}$.**

Similarly to (Zn$_2$L1)$^{3+}$, dinuclear metal complexes with (2,6-bis[(*N*,*N*′-bis[2-picolyl]amino)methyl]-4-tertbutylphenol, L2), such as (Co$_2$L2)$^{3+}$, (Cu$_2$L2)$^{3+}$, (Zn$_2$L2)$^{3+}$, (Ga$_2$L2)$^{5+}$, and (In$_2$L2)$^{5+}$ have been also tested as phosphate capture molecules.[35-38] The molecular structure of (Me$^{n+}$)$_2$(L2)$^{-1}$, where Me$^{n+}$ is Cu$^{2+}$, Zn$^{2+}$ or Ga$^{3+}$, is shown in Scheme 1b. In this study, we tested the applicability of the dimetallic complexes (Cu$_2$L1)$^{3+}$, (Cu$_2$L2)$^{3+}$, (Zn$_2$L2)$^{3+}$, and (Ga$_2$L2)$^{5+}$ for ESI-based ETD-MS$^2$ analysis of phosphopeptides and compared the results with those from the (Zn$_2$L1)$^{3+}$-aided method. As described above, Zn$^{2+}$ and Ga$^{3+}$ are suitable adductive for peptide sequencing by ETD-MS/MS. In consequence, we used Zn$^{2+}$ and Ga$^{3+}$ as the nuclears in the complex and complexes with Cu$^{2+}$ was used as the reference. First, we evaluate the efficiency for the binding of phosphorylated serine (pS) to the dimetallic complexes from ESI mass spectra. We then used the dimetallic complexes as additives for the ETD-MS/MS analysis of tryptic phosphopeptides, T18p (NVPLpYK), T19p (HLADLpSK), and T43p (VNQIGpTLSESIK), which have been used as models in previous studies.[16] Although ESI of these phosphopeptides produced singly and doubly protonated molecules, the triply protonated phosphopeptides were absent. The precise location of the phosphorylation site could not be determined because of the low sequence coverage when doubly protonated molecules were used as the precursors for ETD-MS/MS.[16] In contrast, the use of zinc complexes increased in the charge state of phosphopeptides in ESI-MS and facilitated the phosphopeptide sequencing by subsequent ETD-MS/MS analysis. To address the general mechanism for the ETD of the dimetal-ligand-phosphopeptide complexes, the details of the ETD fragmentation pathway were investigated using density functional theory (DFT) calculations.

## EXPERIMENTS

### *Materials*

Synthetic phosphopeptides designed to mimic the fragments produced by the tryptic digestion of yeast enolases, T18p, T19p, and T43p, were purchased from CS Bio Co., Ltd. (Shanghai, China). The phosphorylated Nα-acetylated serine–lysine dipeptide (Ac-pSK) was purchased from GenScript Inc. (NJ, USA). The detailed information for the model phosphopeptides are summarized in Table 1.

Paraformaldehyde, sodium triacetoxyborohydride, 2-pyridinecarboxaldehyde, 1,3-diamino-2-propanol, copper (II) chloride, zinc (II) chloride, and gallium (III) perchlorate were supplied by Sigma-Aldrich Inc. (MO, USA). Bis(2-pyridylmethyl)amine and *p*-cresol were obtained from Tokyo Chemical Industry Co., Ltd. (Tokyo, Japan).

All solvents used were of high-performance liquid chromatography grade, except for water, which was purified by a Milli-Q® purification system (Millipore; Billerica, MA, USA). The conductivity of the water was 18.2 MΩ/cm.

**Table 1. Monoisotopic Mass ($M_m$), Sequence, and Composition of the Analyte Phosphopeptides Used**

| Analyte | $M_m$ | Sequence | Composition |
|---------|-------|----------|-------------|
| pS | 185.08 | pS | $C_3H_8NO_6P$ |
| Ac-pSK | 346.11 | Ac-pSK | $C_{11}H_{23}N_3O_8P$ |
| T18p | 812.39 | NVPLpYK | $C_{35}H_{57}O_{12}N_8P$ |
| T19p | 862.40 | HLADLpSK | $C_{34}H_{59}O_{14}N_{10}P$ |
| T43p | 1367.68 | VNQIGpTLSESIK | $C_{55}H_{98}O_{23}N_{15}P$ |

*Synthesis of ligand molecules*

Ligand L1 was synthesized from 1,3-diamino-2-propanol and 2-pyridinecarboxaldehyde by dehydration condensation in methanol. To facilitate the reaction, sodium triacetoxyborohydride was added. The final concentrations of 1,3-diamino-2-propanol, 2-pyridinecarboxaldehyde, and sodium triacetoxyborohydride were approximately 0.15, 0.75, and 0.84 M, respectively. The reaction was performed for 5 days at room temperature. Ligand L2 was synthesized from *p*-cresol, *p*-formaldehyde, and bis(2-pyridylmethy)amine in ethanol using a literature method.[39] The final concentration of *p*-cresol, *p*-formaldehyde, and bis(2-pyridylmethy)amine were approximately 0.35, 0.8, and 0.8 M, respectively, and the reaction was performed under reflux conditions over 3 days. Synthesized ligands L1 and L2 were purified by column chromatography with silica gel and recrystallization, respectively.

To obtain dinuclear metal complexes, the synthesized ligands, L1 and L2, were dissolved in methanol at 0.15 M, and the obtained solutions were mixed with aqueous solutions of the corresponding metal salt, i.e., $CuCl_2$, $ZnCl_2$, or $Ga(ClO_4)_3$. The mixing molar ratio of ligand/metal was approximately 1/2. The complexes were obtained by re-crystallization.

***Measurement of Binding Affinity between pS and $(Me^{n+})_2(Lx^{-1})$ by ESI-MS***

ESI-MS was used to investigate the trend of binding affinities between pS and $(Me^{n+})_2(Lx^{-1})$, where $Me^{n+}$ is $Zn^{2+}$, $Cu^{2+}$, or $Ga^{3+}$, and $Lx$ is L1 or L2. The abundances of free $(Me^{n+})_2(Lx^{-1})$ and $(Me^{n+})_2(Lx^{-1})(pS)$ complexes were estimated by positive ion ESI-MS using an Orbitrap Exactive mass spectrometer (ThermoFisher, Bremen, Germany). The dinuclear complexes, $(Me^{n+})_2(Lx^{-1})$, were dissolved in water/methanol (1/1, v/v) at a concentration of 20 µM. The complexation was studied in solutions containing 10 µM $(Me^{n+})_2(Lx^{-1})(pS)$ and variable concentrations of pS (from 0.5 to 100 µM). The final concentration of the dinuclear complex was adjusted to 10 µM for all experiments. The values for the abundance of $(Me^{n+})_2(Lx^{-1})$ and $(Me^{n+})_2(Lx^{-1})(pS)$ were estimated from the peak areas of the corresponding signals observed in the ESI mass spectra and the mass balance equation on the total concentration of $[(Me^{n+})_2(Lx^{-1})]$.

***ETD-MS/MS***

The analyte peptides were dissolved in water/methanol (1/1, v/v) at a concentration of 10 µM. To produce the dinuclear complex-phosphopeptide by ESI, $(Cu_2Lx)^{3+}$, $(Zn_2Lx)^{3+}$, or $(Ga_2L2)^{5+}$ were added to the peptide solution at concentrations of 100, 40, and 20 µM, respectively. ESI mass spectra were acquired using a 9.4-T Fourier transform ion cyclotron resonance (FT-ICR) mass spectrometer (SolariX FT, Bruker, Germany). The analyte solution was directly infused into the mass spectrometer using ESI. The ion accumulation time, ion cooling time, and time-of-flight values were set to 0.5 s, 20 ms, and 7 ms, respectively. For the ETD-MS/MS experiments, the precursor ions were mass-selected in the quadrupole filter and then reacted with the fluoranthene radical anion. The times for reagent accumulation and the ion/ion reaction were set to 150 and 25 ms, respectively. Total ESI-MS and subsequent ETD-MS/MS mass spectra were obtained by the accumulation of 20 and 100 single mass spectra, respectively.

***Computational details***

All electronic structure calculations were performed with Gaussian 16.[40] The geometries for the zinc trihistidine complexes were optimized by DFT calculations using the MN15[41] hybrid functional and double-zeta valence polarized basis sets, i.e., LanL2DZ for the metal atoms and 6-31G(d) for the C, H, O, N, and P atoms. To establish the fragmentation energies, transition state (TS) geometries were also optimized at the MN15/LanL2DZ/6-31G(d) level of theory and confirmed by vibrational frequency analysis to have one imaginary frequency. The relationship between the transition state and the reactants, as well as the intermediates, was checked by intrinsic reaction coordinate analysis[42] starting from the transition state conformation. Single-point energies of the local energy minima and transition state geometries were calculated using the MN15 functional with the 6-31++G(2d,p) basis set. Excited electronic states were calculated using a time-dependent DFT method[43] with the MN15 functional and the 6-31++G(2d,p) basis set.

*Notation*

In the present study, Zubarev's notation was adopted for peptide fragment ions.[44] According to this notation, homolytic N–$C_\alpha$ bond cleavage yields the radical $c\bullet$ and $z\bullet$ fragments, and the addition of a hydrogen atom to the $c\bullet$ or $z\bullet$ fragments produces a $c'$ or $z'$ fragment, respectively. The abstraction of a hydrogen atom from the $c\bullet$ or $z\bullet$ fragments produces a $c$ or $z$ fragment, respectively.

# RESULTS AND DISCUSSION

## Formation Efficiency of Dinuclear Metal-pS Complexes

First, we focused on the binding properties of dimetallic complexes to pS, which were estimated from the ESI mass spectra. In this study, five dimetallic complexes, $(Zn_2L1)^{3+}$, $(Zn_2L2)^{3+}$, $(Cu_2L1)^{3+}$, $(Cu_2L2)^{3+}$ and $(Ga_2L2)^{5+}$ were tested. Because the utility of the $(Zn_2L1)^{2+}$-aided ETD-MS/MS for phosphopeptide sequencing were described by previous report,[16] we first discussed on the complex formation between $(Zn_2L1)^{3+}$ and pS. Figure 1a shows the ESI mass spectra of the mixture of $(Zn_2L1)^{3+}$ and pS in different mixing ratios.

As described in experimental section, $(Zn_2L1)^{3+}$ was synthesized from [L1 + H] and $ZnCl_2$. Thereby, $Cl^-$ would be present as counter ion of $(Zn_2L1)^{3+}$, and $(Zn_2L1)^{3+}$ was detected as the complex with $Cl^-$, $[Zn_2L1 + Cl]^{2+}$, when the concentration of pS was lower than that of $(Zn_2L1)^{3+}$ (upper panel of Figure 1a). In addition, $[ZnL1 + H]^{2+}$ and $[ZnL1 - C_6H_6N]^{2+}$ appeared in ESI mass spectrum. These ions might be produced by ion-solvent reaction during ESI process. The proposed structures of $[Zn_2L1 + Cl]^{2+}$ and $[ZnL1 + H]^{2+}$ are described in Scheme 2.

When concentration of pS increased, the intensity of $[Zn_2L1 + Cl]^{2+}$, $[ZnL1 + H]^{2+}$ and $[ZnL1 - C_6H_6N]^{2+}$ decreased and the intensity of the complex between $(Zn_2L1)^{3+}$ and pS increased (lower panel of Figure 1a). The complex between $(Zn_2L1)^{3+}$ and pS was detected as doubly-charged form, $[Zn_2L1 + pS - H]^{2+}$. Because $(Zn_2L1)^{3+}$ selectively binds to the phosphate group,[16] positive charges in $[Zn_2L1 + pS - H]^{2+}$ would be located at the $(Zn_2L1)^{3+}$-adducted phosphate group and amino group, which is the most probable site for protonation. The proposed structure of $[Zn_2L1 + pS - H]^{2+}$ are shown in Scheme 2. In addition, $[Zn_2L1 + H_2PO_4]^{2+}$ was observed in ESI mass spectra and the intensity of $[Zn_2L1 + H_2PO_4]^{2+}$ was increased with increasing of pS concentration. Thus, $[Zn_2L1 + H_2PO_4]^{2+}$ was suggested to be originated from the $(Zn_2L1)^{3+}$ and pS complex, as shown in Scheme 2.

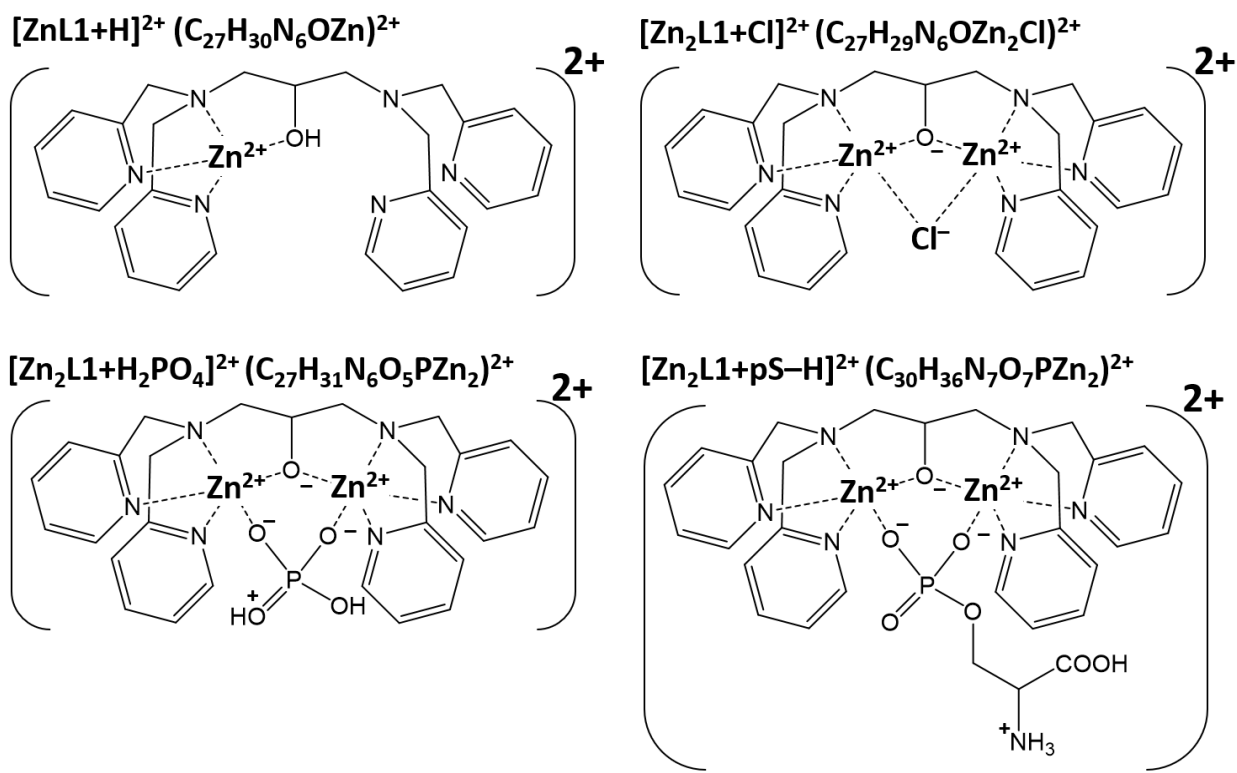

**Scheme 2. The proposed structure of ions observed in ESI mass spectra of (Zn₂L1)³⁺ and pS mixtures, as shown in Figure 1a.**

Next, we focused on the complexation efficiency of $(Zn_2L1)^{3+}$ binding to pS. To estimate the complexation efficiency, we performed ESI-MS of a $(Zn_2L1)^{3+}$ and pS mixture at a range of pS concentrations from 0.5 to 40 µM. The concentration of $(Zn_2L1)^{3+}$ was set to 10 µM for all experiments. Figure 1b shows the relationship between the signal intensity ratio of $(Zn_2L1)^{3+}$ with and without pS, and the total pS concentration between 0.5 and 40 µM. The values of the signal ratio were calculated from the sum of the intensities of all species assigned in Figure 1a. The yield of the complex between $(Zn_2L1)^{3+}$ and pS increased with an increasing concentration of pS and 60 % of $(Zn_2L1)^{3+}$ was bound to pS, when the concentration of $(Zn_2L1)^{3+}$ and pS were 10 µM.

As in the case of $(Zn_2L1)^{3+}$, $(Zn_2L2)^{3+}$ selectively bound to the phosphate group of pS, producing $[Zn_2L2 + pS - H]^{2+}$. The relationship between the amount of $(Zn_2L1)^{3+}$ with and without pS at different concentrations of pS is shown in Figure 1c. The comparison of Figures 1b and 1c indicated the yield of complex formation between $(Zn_2L2)^{3+}$ and pS was slightly lower than that of $(Zn_2L1)^{3+}$ and pS. In consequence, the $(Zn_2L1)^{3+}$ would capture the phosphate group more efficiently.

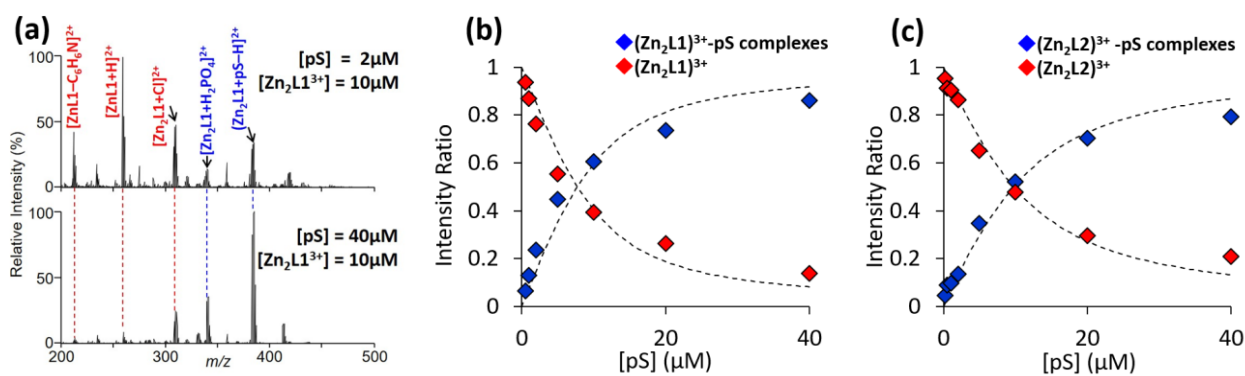

**Figure 1. (a) ESI mass spectra of $(Zn_2L1)^{3+}$ and pS mixtures with different concentration ratios. (b, c) The relationship between the ESI-MS data based on the concentration of pS to the yield of (b) $(Zn_2L1)^{3+}$-pS and (c) $(Zn_2L2)^{3+}$-pS complexes.**

Next, we investigated dinuclear copper complexes, $(Cu_2L1)^{3+}$ and $(Cu_2L2)^{3+}$, for their use as phosphate capture molecules. Figure 2a shows the ESI mass spectra for the mixture of 10 μM $(Cu_2L1)^{3+}$ and pS with mixing ratios, $[(Cu_2L1)^{3+}]/[pS]$, of 1 and 6. As in the case of $(Zn_2L1)^{3+}$, the ESI mass spectra of the mixture of $(Cu_2L1)^{3+}$ and pS contained chloride and pS adduct forms, $[Cu_2L1+Cl]^{2+}$ and $[Cu_2L1 + pS - H]^{2+}$. In contrast, the fragment ions $[CuL1 + H]^{2+}$ and $[Cu_2L1 + H_2PO_4]^{2+}$ were absent. In terms of relative affinities, although $(Zn_2L1)^{3+}$ was mostly bound to pS when 40 μM of pS was added to the $(Zn_2L1)^{3+}$ solution (Figure 1a), the chloride adduct of $(Cu_2L1)^{3+}$ was detected as an intense signal at 40 μM of added pS (Figure 2a). $(Cu_2L2)^{3+}$ showed similar behavior to $(Cu_2L1)^{3+}$ (data not shown). Next, the formation of the $(Cu_2L1)^{3+}$-pS and $(Cu_2L2)^{3+}$-pS complexes were determined using the same procedure as for the $(Zn_2L1)^{3+}$-pS complex (Figures 2b and 2c). The comparison of Figures 1 and 2 indicate that $(Cu_2L1)^{3+}$ and $(Cu_2L2)^{3+}$ have lower phosphate capturing ability than $(Zn_2L1)^{3+}$ and $(Zn_2L2)^{3+}$. Thus, the phosphate capturing ability of dinuclear metal complex is strongly affected by the type of metal ion.

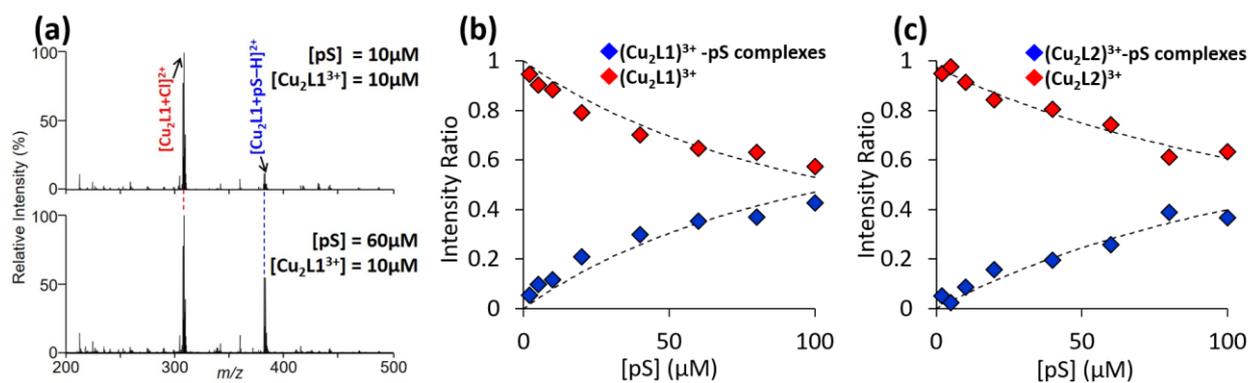

**Figure 2.** (a) ESI mass spectra of $(Cu_2L1)^{3+}$ and pS mixtures with different concentration ratios. (b, c) The relationship between the ESI-MS data based on the concentration of pS to the yield of (b) $(Cu_2L1)^{3+}$-pS and (c) $(Cu_2L2)^{3+}$-pS complexes.

Next, $(Ga_2L2)^{5+}$ was used as a phosphate capture molecule. Because Ga is a trivalent metal ion, the use of $(Ga_2L2)^{5+}$ is expected to increase the charge state of the phosphorylated peptides. Figure 3a shows the ESI mass spectra for the mixture of 10 μM $(Ga_2L2)^{5+}$ and pS with mixing ratios, $[pS]/[(Ga_2L2)^{5+}]$, of 0.2 and 1. When $[pS]/[(Ga_2L2)^{5+}]$ was 0.2, $(Ga_2L2)^{5+}$ was detected as $[Ga_2L2 + O]^{3+}$, $[Ga_2L2 + O + OH]^{2+}$ and $[Ga_2L2 + O + ClO_4]^{2+}$. The source of $ClO_4^-$ is the $Ga(ClO_4)_3$ used for the synthesis of $(Ga_2L2)^{5+}$. According to the previous reports, $(Ga_2L2)^{5+}$ is estimated to be present as $(Ga_2L2)(OH)_2(ClO_4)_3(H_2O)_2$ in solid state.[38] Because $OH^-$ and $ClO_4^-$ were presented as the adduct ions in ESI mass spectra, the counter ions of $(Ga_2L2)^{5+}$ in solution are suggested as $OH^-$ and $ClO_4^-$. In consequence, $[Ga_2L2 + O]^{3+}$ might be formed from $[Ga_2L2 + 2OH]^{3+}$ during ESI process by chemical reaction (1) and proposed structure of $[Ga_2L2+O]^{3+}$ was shown in Scheme 3.

$$[Ga_2L2 + 2OH]^{3+} \;\rightarrow\; [Ga_2L2 + O]^{3+} + H_2O \qquad\qquad (1)$$

To investigate the reaction behaviors of $[Ga_2L2 + 2OH]^{3+}$ in details, we calculated optimized geometries by DFT with MN15/6-31++G(2d,p)//MN15/LanL2DZ/6-31G(d) level (Supporting information, Scheme S1). In the lowest energy geometry of $[Ga_2L2 + 2OH]^{3+}$, a hydroxide was strongly coordinated by both $Ga^{3+}$ and another hydroxide weakly bound to a $Ga^{3+}$. The proton transfer reaction between hydroxides produced the intermediate complex, IM, consisted by $[Ga_2L2 + O]^{3+}$ and $H_2O$. The corresponding transition state was TS in Scheme S1, which is 30 kJ/mol less stable than reactant. The complete dissociation energy of $[Ga_2L2 + O]^{3+}$ and $H_2O$ from $[Ga_2L2 + 2OH]^{3+}$ was 66 kJ/mol. Because the internal energies of the ions produced by ESI with standard parameter sets is estimated to be around 180 kJ/mol,[45] the proposed formation pathway of $[Ga_2L2 + O]^{3+}$ is feasible.

The addition of pS contributes to increasing the signal intensity of the $(Ga_2L2)^{5+}$-pS complex, $[Ga_2L2 + pS − 3H]^{2+}$ (lower panel of Figure 3a). $(Ga_2L2)^{5+}$ would coordinate to a deprotonated phosphate group and a carboxylic acid group to form $[Ga_2L2 + pS − 3H]^{2+}$. Figure 3b shows the relationship between the amount of $(Ga_2L2)^{5+}$ for a total pS concentration in the range of 0.5−40 μM. The amount of $(Ga_2L2)^{3+}$ with and without pS was estimated from the peak areas of $[Ga_2L2 + O]^{3+}$, $[Ga_2L2 + O + OH]^{2+}$, $[Ga_2L2 + O + ClO_4]^{2+}$, and $[Ga_2L2 + pS − 3H]^{2+}$. $(Ga_2L2)^{5+}$ was mostly bound to pS when 20 μM of pS was added to the 10 μM $(Ga_2L2)^{5+}$ solution. The comparison of Figures 1-3 indicates that $(Ga_2L2)^{5+}$ has the highest binding ability for pS among the complexes tested in this study.

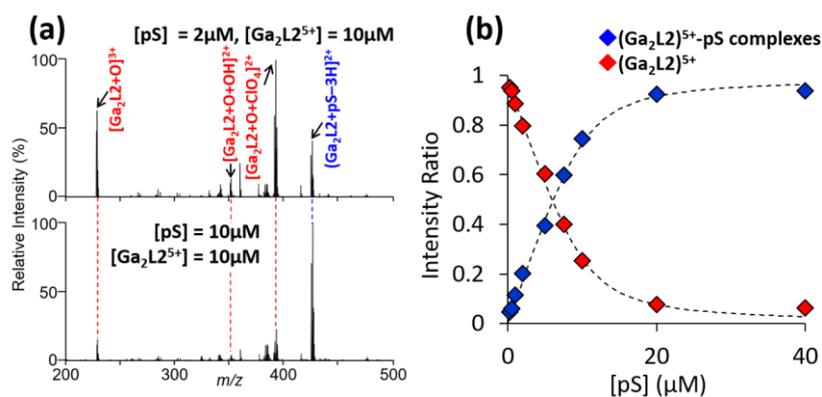

**Figure 3.** (a) ESI mass spectra of $(Ga_2L2)^{5+}$ and pS mixtures with different concentration ratios. (b) The relationship between the ESI-MS data based on the concentration of pS to the yield of the $(Ga_2L1)^{5+}$-pS complex.

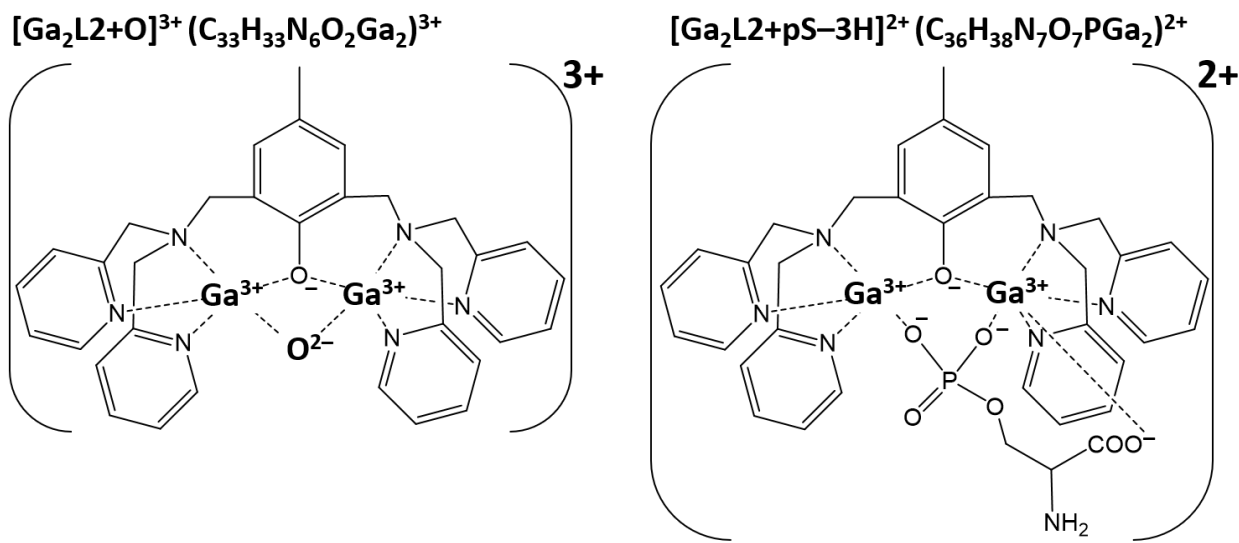

**Scheme 3. The structures and formation processes of ions observed in the ESI mass spectra of the (Ga₂L1)⁵⁺ and pS mixture, as shown in Figure 2a.**

*ETD-MS² of the Dinuclear Metal-Phosphopeptide Complexes*

We also investigated the suitability of the complexes for phosphopeptide sequencing by ETD-MS/MS. The phosphopeptide T18p was used as the model peptide and triply-charged dinuclear metal-T18p complexes were used as the precursor ions for ETD-MS/MS experiments. Figure 4 shows the ETD-MS/MS spectra of [Zn₂L1 + T18p]³⁺, [Zn₂L2 + T18p]³⁺, [Cu₂L1 + T18p]³⁺, and [Ga₂L2 + (T18p – 2H)]³⁺. The (Zn₂L1)³⁺ complex was used in our previous work as an additive to increase the charge state of phosphorylated peptides, facilitating phosphopeptide sequencing by ETD-MS/MS.[16] As described in the introduction, the ETD of the (Zn₂L1)³⁺-phosphopeptide complex produces an intense signal of the fragment ion arising from a 92-Da loss, which corresponds to ligand degradation. The presence of an intense signal corresponding to the fragment ion because of the 92-Da loss hinders phosphopeptide sequencing. As in previous reports,[16] the ETD of [Zn₂L1 + (Ac-pSK – H)]²⁺ showed the fragment ion arising from the 92-Da loss as an intense signal, in addition of formation of the fragments due to N–Cα bond cleavage. In contrast, the ETD-MS/MS mass spectrum of [Zn₂L2 + T18p]³⁺ showed a more intense signal corresponding to fragment ions arising from N–Cα bond cleavage, and a lower intensity of the fragment ion arising from the 92-Da loss, as compared to when $(Zn_2L1)^{3+}$ was used.

As shown in Figure 4c, the ETD of $[Cu_2L1 + T18p]^{3+}$, generated charge reduced products, $[Cu_2L1 + T18p]^{2+}$, and $[Cu_2L1 + T18p]^+$. In addition, $(CuL1)^+$ and $[T18p + Cu]^+$ were observed as the fragment ion. Importantly, the $[T18p + Cu]^+$ contained monovalent copper atom, indicating that ETD induced reduction of $Cu^{2+}$. Because the binding energy between $(Cu_2L1)^{3+}$ and T18p was decreased by reduction of $Cu^{2+}$ in the complex, $(CuL1)^+$ and $[T18p + Cu]^+$ would be formed by ETD-MS/MS. This result is good agreement with previously reported ETD-MS/MS of $Cu^{2+}$-peptide complex[30] and the presence of $Cu^{2+}$ in the precursor ion suppress the N–Cα bond cleavage by ETD-MS/MS.

Regarding the Figure 4d, ETD of $[Ga_2L2 + (T18p - 2H)]^{3+}$ selectively induced $C_6H_6N\bullet$ losses, leading to the $[Ga_2L2 + (T18p - 2H - C_6H_6N)]^{2+}$ and $[Ga_2L2 + (T18p - 2H - 2C_6H_6N)]^+$. The $C_6H_6N\bullet$ loss was induced the electron association to the pyridine ring in the complex, and the N–Cα bond cleavage is completely suppressed by the presence of $(Ga_2L2)^{5+}$ in the precursor.

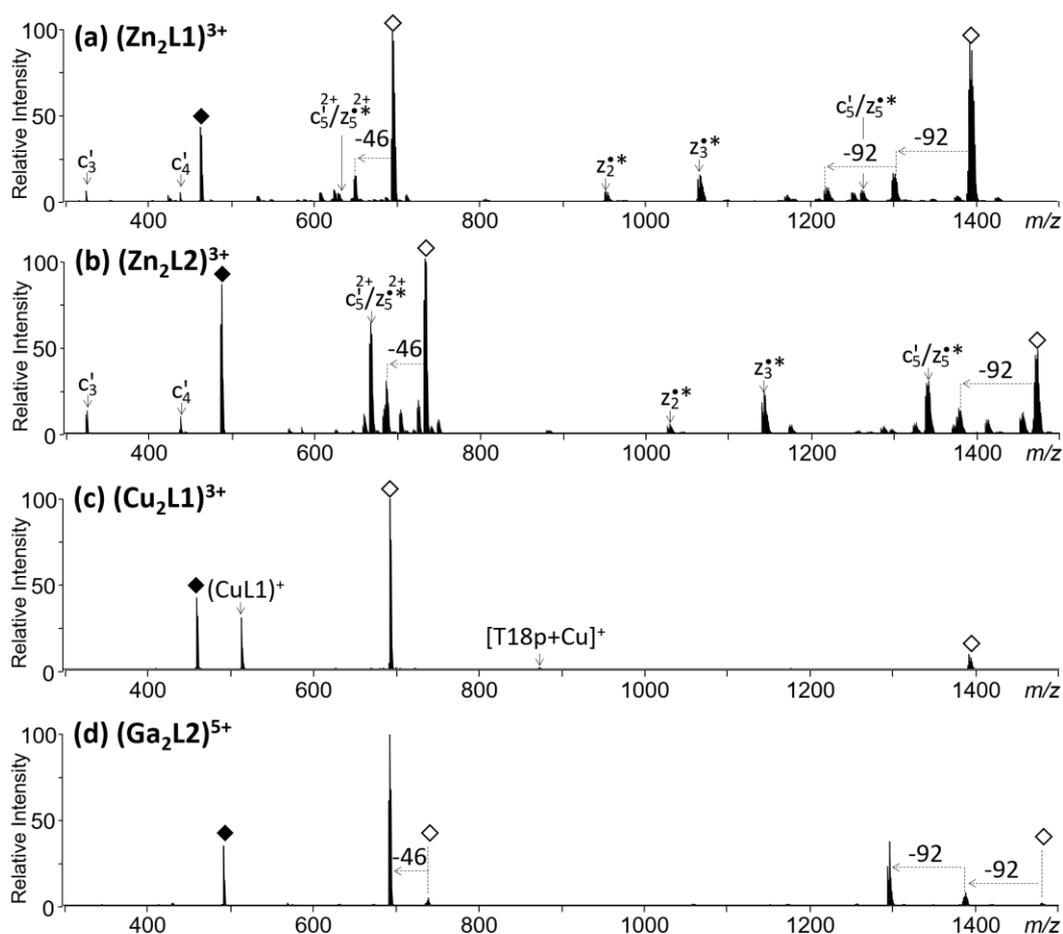

**Figure 4. ETD-MS/MS mass spectra of the doubly charged complexes: (a) [Zn₂L1 + T18p]³⁺, (b) [Zn₂L2 + T18p]³⁺, (c) [Cu₂L1 + T18p]³⁺ and (d) [Ga₂L2 + (T18p − 2H)]³⁺. Asterisks and black squares indicate the precursor ions and charge reduced products, respectively. "−92" is the unwanted ligand degradation.**

To investigate of the detailed ETD process of dinuclear metal-T18p complexes, a small peptide, Ac-pSK, was used as the model peptide for ETD-MS/MS experiments and DFT calculation. ESI of Ac-pSK and the dinuclear metal complexes mixture produced doubly charged complexes, which were used as the precursor ions for ETD-MS/MS experiments. Figure 5 shows the ETD-MS/MS spectra of [Zn₂L1 + (Ac-pSK − H)]²⁺, [Zn₂L2 + (Ac-pSK − H)]²⁺, [Cu₂L1 + (Ac-pSK − H)]²⁺, and [Ga₂L2 + (Ac-pSK − 3H)]²⁺. First, we focused on ETD-MS/MS spectra of [Zn₂L1 + (Ac-pSK − H)]²⁺ and [Zn₂L2 + (Ac-pSK − H)]²⁺.

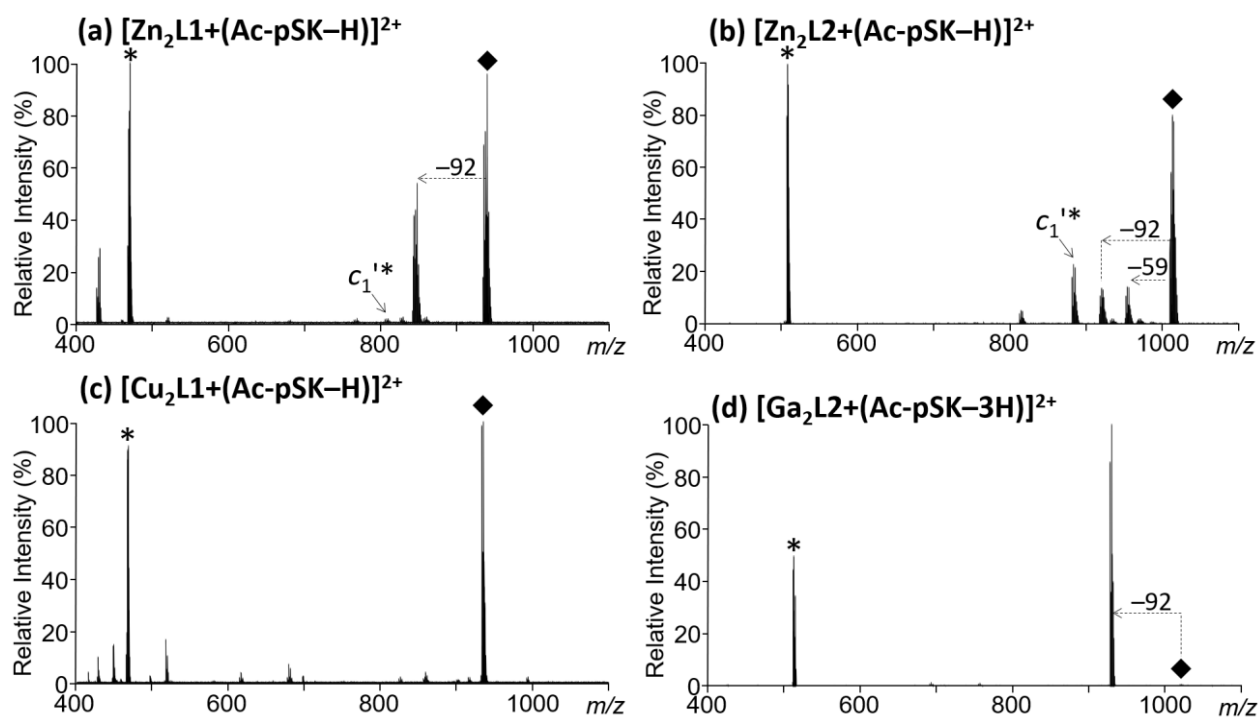

**Figure 5. ETD-MS/MS mass spectra of the doubly charged complexes: (a) [Zn$_2$L1 + (Ac-pSK − H)]$^{2+}$, (b) [Zn$_2$L2 + (Ac-pSK − H)]$^{2+}$, (c) [Cu$_2$L1 + (Ac-pSK − H)]$^{2+}$, and (d) [Ga$_2$L2 + (Ac-pSK − 3H)]$^{2+}$. Asterisks and black squares indicate the precursor ions and charge reduced products, respectively. "−92" is the unwanted ligand degradation, whereas "$c_1$'" and "−59" are the desired N–C$\alpha$ fragmentation.**

For comparison of (Zn$_2$L1)$^{3+}$- and (Zn$_2$L2)$^{3+}$-aided methods, ETD mass spectra of [Zn$_2$L1 + (Ac-pSK − H)]$^{2+}$ and [Zn$_2$L2 + (Ac-pSK − H)]$^{2+}$ showed both of fragments due to 92-Da loss and N–C$\alpha$ bond cleavage, whereas the yields are different from [Zn$_2$L1 + (Ac-pSK − H)]$^{2+}$ and [Zn$_2$L2 + (Ac-pSK − H)]$^{2+}$, as in the case of [Zn$_2$L1 + T18]$^{3+}$ and [Zn$_2$L2 + T18]$^{3+}$. The potential fragmentation pathways induced by ETD are summarized in Scheme 4. Although the loss of acetyl groups through an intermediate, R3, is expected to occur during the ETD process of [Zn$_2$L1 + (Ac-pSK − H)]$^{2+}$, the corresponding signal for [Zn$_2$L1 + (Ac-pSK − H) − 59]$^{+\bullet}$ was absent, as can be seen in Figure 5a, indicating that N–C$\alpha$ bond cleavage is a low-efficiency process.

The ETD-MS/MS mass spectrum of $[Zn_2L2 + (Ac\text{-}pSK - H)]^{2+}$ showed a more intense signal corresponding to fragment ions arising from N–Cα bond cleavage ($c_1$'), and a lower intensity of the fragment ion arising from the 92-Da loss, as compared to when $(Zn_2L1)^{3+}$ was used. To explain these differences, the ETD processes of $[Zn_2L1 + (Ac\text{-}pSK - H)]^{2+}$ and $[Zn_2L2 + (Ac\text{-}pSK - H)]^{2+}$ were investigated by DFT. Regarding the initial ETD step, electron association occurs either at the pyridine ring in the ligand, protonated amino group, or π* antibonding orbital of the peptide bond. The electron association to the pyridine ring in the doubly charged complex and the subsequent geometry relaxation generated zwitterionic radical R1 (shown in Scheme 4), whereas R2 and R3 were generated by electron association to the π* antibonding orbital of the peptide bond and protonated amino group, respectively. The detailed reaction energies, including transition state barrier (TS) and intermediate (IM) for the ETD fragmentation of $[Zn_2L1 + (Ac\text{-}pSK - H)]^{2+}$ and $[Zn_2L2 + (Ac\text{-}pSK - H)]^{2+}$ are summarized in Table 2 and the corresponding geometries are shown in Supporting Information, Scheme S2. Although $[Zn_2L1 + (Ac\text{-}pSK - H) - 59]^{+\bullet}$ is the most intense fragment ion in Figure 4a, the transition state barrier for $C_6H_6N\bullet$ loss from R1 is 121 kJ/mol, which is higher than that for the N–Cα bond cleavage of R2 and R3. Additionally, the change from $(Zn_2L1)^{3+}$ to $(Zn_2L2)^{3+}$ significantly suppressed the formation of the fragment arising from a 92-Da loss, whereas R1' has a lower transition state barrier for $C_6H_6N\bullet$ loss than R1. The DFT calculations reveal that the yield of the fragment ion does not depend on the transition state energy for fragmentation. The fragments arising from N−Cα bond cleavage and $C_6H_6N\bullet$ loss originated from the different electronic states of the charge-reduced complexes. In other words, the intermediate radicals, R1, R2, and R3, are formed by these different processes. Therefore, the yields of $[Zn_2L1 + (Ac\text{-}pSK - H) - 92]^{+}$, $[Zn_2L1 + (c'_1 - 2H)]^{+}$, and $[Zn_2L1 + (Ac\text{-}pSK - H) - 59]^{+\bullet}$ are expected to reflect the amounts of R1, R2, and R3, respectively.

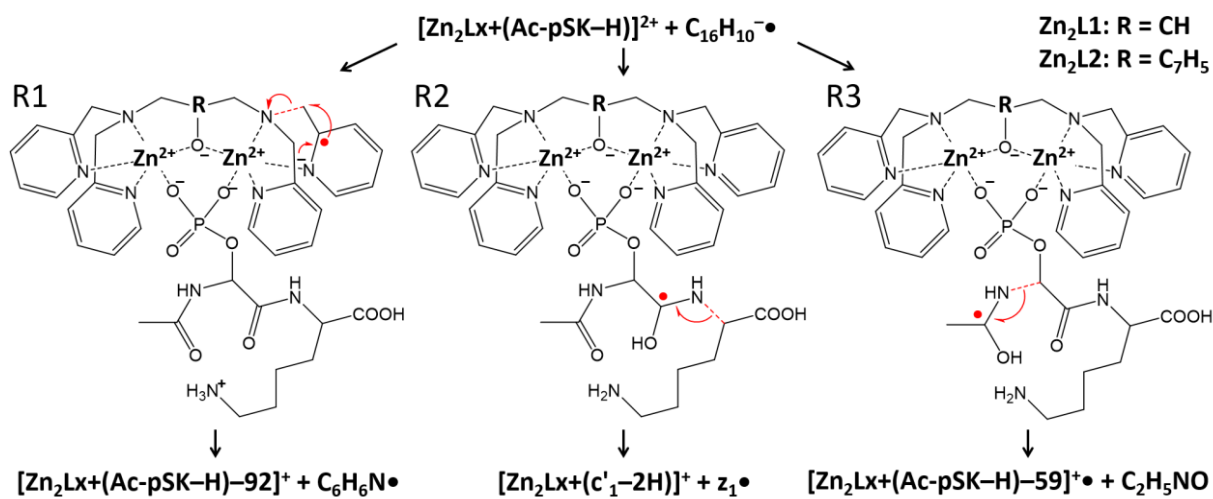

**Scheme 4.** ETD mechanism of $[Zn_2L1 + (Ac\text{-}pSK − H)]^{2+}$ and $[Zn_2L2 + (Ac\text{-}pSK − H)]^{2+}$. The detailed information including transition state geometries for fragmentation were shown in Supporting Information, Scheme S2.

**Table 2. Reaction energies of ETD-induced fragmentation, as shown in Scheme 4. The relative energies (kJ/mol) were obtained from MN15/6-31++G(2d,p)//MN15/LanL2DZ/6-31G(d) calculations.**

| Reaction | TS | IM | Products |
|---|---|---|---|
| R1 $\rightarrow$ $[Zn_2L1 + (Ac\text{-}pSK − H) − 92]^+ + C_6H_6N\bullet$ | 121 | 116 | 213 |
| R1' $\rightarrow$ $[Zn_2L2+(Ac\text{-}pSK−H)−92]^+ + C_6H_6N\bullet$ | 77 | 27 | 156 |
| R2 $\rightarrow$ $[Zn_2L1 + (c'_1 − 2H)]^+ + z_1\bullet$ | 20 | −42 | 139 |
| R2' $\rightarrow$ $[Zn_2L2 + (c'_1 − 2H)]^+ + z_1\bullet$ | 30 | −67 | 143 |
| R3 $\rightarrow$ $[Zn_2L1 + (Ac\text{-}pSK − H) − 59]^+\bullet + C_2H_5NO$ | 41 | −33 | 27 |
| R3' $\rightarrow$ $[Zn_2L2 + (Ac\text{-}pSK–H) − 59]^+\bullet + C_2H_5NO$ | 42 | −35 | 24 |

To investigate the intermediate radicals formed (R1, R2, and R3), we focused on the initial step of ETD, *i.e.*, the electron attachment process of doubly charged complexes, which was modeled using time-dependent (TD) DFT calculations. Figures 5a and 5b show the molecular orbitals for the ground state and excited states of $[Zn_2L1 + (Ac\text{-}pSK - H)]^{+\bullet}$ and $[Zn_2L2 + (Ac\text{-}pSK - H)]^{+\bullet}$, respectively, and their corresponding vertical excitation energies. Although electron transfer process does not occur in a monochromatic way, the calculated vertical excitation energies would relate the relative energies for the excited state of the $[Zn_2L1 + (Ac\text{-}pSK - H)]^{+\bullet}$ and $[Zn_2L2 + (Ac\text{-}pSK - H)]^{+\bullet}$, which were formed by ETD. The ground state and excited states, R1a–R1e in Figure 6a and R1a'–R1e' in Figure 6b, lead to the zwitterionic radical R1 in Scheme 4, which has a pyridine radical anion. The intermediate radicals R1 and R1' then underwent $C_6H_6N\bullet$ loss, forming $[Zn_2L1 + (Ac\text{-}pSK - H) - 92]^{+\bullet}$ and $[Zn_2L2 + (Ac\text{-}pSK - H) - 92]^{+\bullet}$, respectively. In contrast, $[Zn_2L1 + (c'_1 - 2H)]^+$ and $[Zn_2L1 + (Ac\text{-}pSK - H) - 59]^{+\bullet}$ would be formed from excited state configurations R2/R2' and R3/R3' in Figure 6, respectively. The electron attachment to excited states R2/R2' and R3/R3' produced an aminoketyl anion radical and ammonium radical, respectively. Subsequently, these radicals were involved in intermolecular hydrogen transfer to give aminoketyl radical intermediates, R2/R2' and R3/R3' in Scheme 4. The aminoketyl radical intermediates immediately underwent N−Cα bond cleavage, as shown in Scheme 4. By comparing Figures 5a and 5b, the relative energies of excited states R2' and R3' in $[Zn_2L2 + (Ac\text{-}pSK - H)]^{+\bullet}$ can be seen to be lower than R2 and R3 in $[Zn_2L1 + (Ac\text{-}pSK - H)]^{+\bullet}$. Therefore, the $[Zn_2L2 + (Ac\text{-}pSK - H)]^{2+\bullet}$ complex can produce aminoketyl radical intermediates R1' and R2' more efficiently than $[Zn_2L1 + (Ac\text{-}pSK - H)]^{2+\bullet}$.

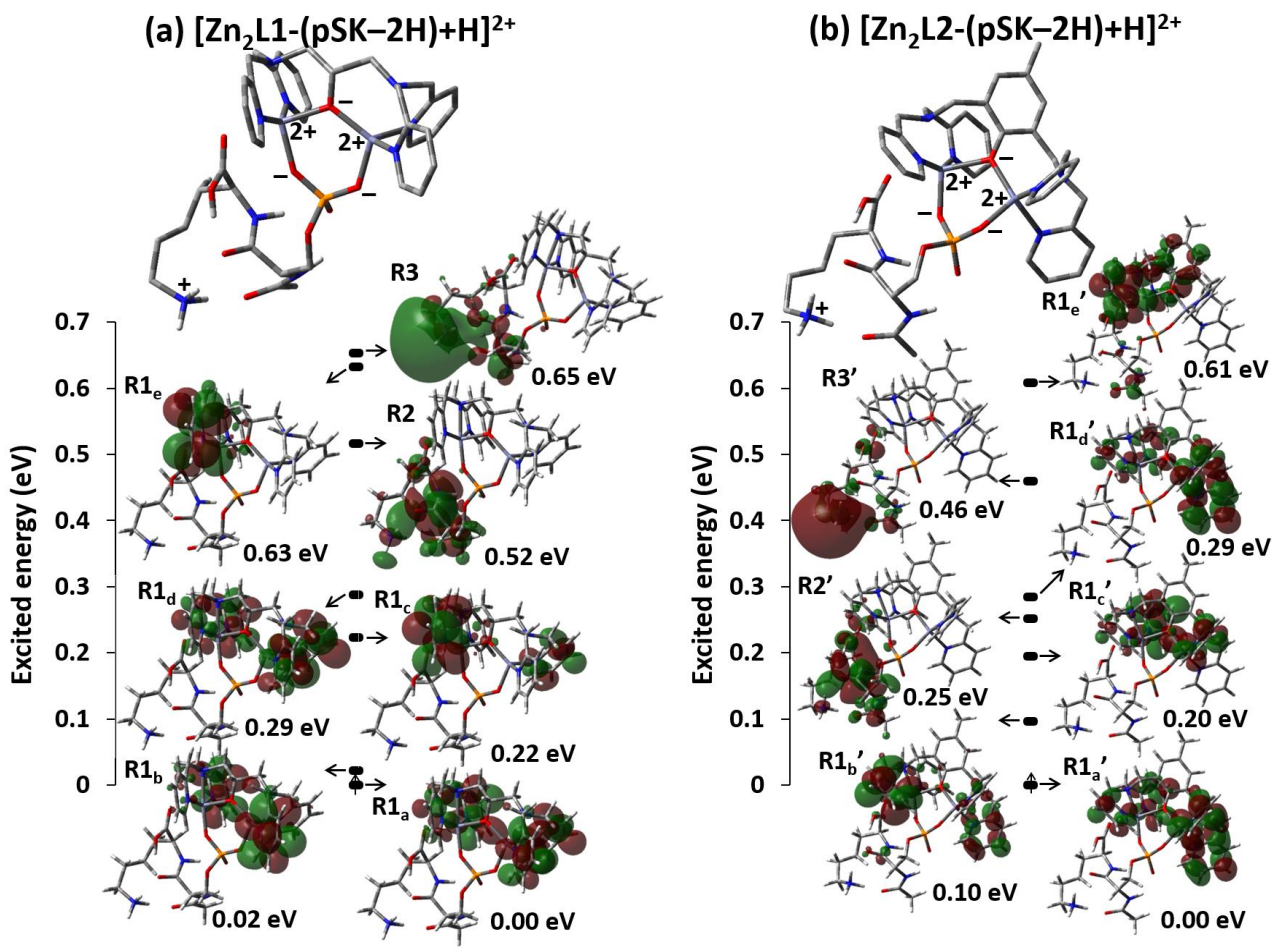

**Figure 6.** Optimized conformations and electronic state diagrams for vertical electron attachment of the doubly charged complexes (a) $[Zn_2L1 + Ac\text{-}pSK − H]^{2+}$ and (b) $[Zn_2L2 + Ac\text{-}pSK − H]^{2+}$. The optimized conformations were obtained at the MN15/LanL2DZ/6-31G(d) level of theory, and the excitation energies (eV) were obtained from TD-MN15/6-31++G(2d,p) calculations.

In contrast to the complexes of $(Zn_2L1)^{3+}$ and $(Zn_2L2)^{3+}$, the ETD of $[Cu_2L1 + (Ac\text{-}pSK − H)]^{2+}$ generated products involving charge reduction, as in the case of $[Cu_2L1 + T18p]^{3+}$. The ETD of $[Cu_2L2 + (Ac\text{-}pSK − H)]^{2+}$ showed a similar result (data not shown). As expected from the experimental results, the DFT calculations indicated that the electron association of $[Cu_2L1 + (Ac\text{-}pSK − H)]^{2+}$ and $[Cu_2L2 + (Ac\text{-}pSK − H)]^{2+}$ produced charge reduced complexes, which contain a monovalent copper cation, $Cu^+$ (scheme 5a) and the formation of aminoketyl radical is largely suppressed by the presence of $Cu^{2+}$ in the precursor.

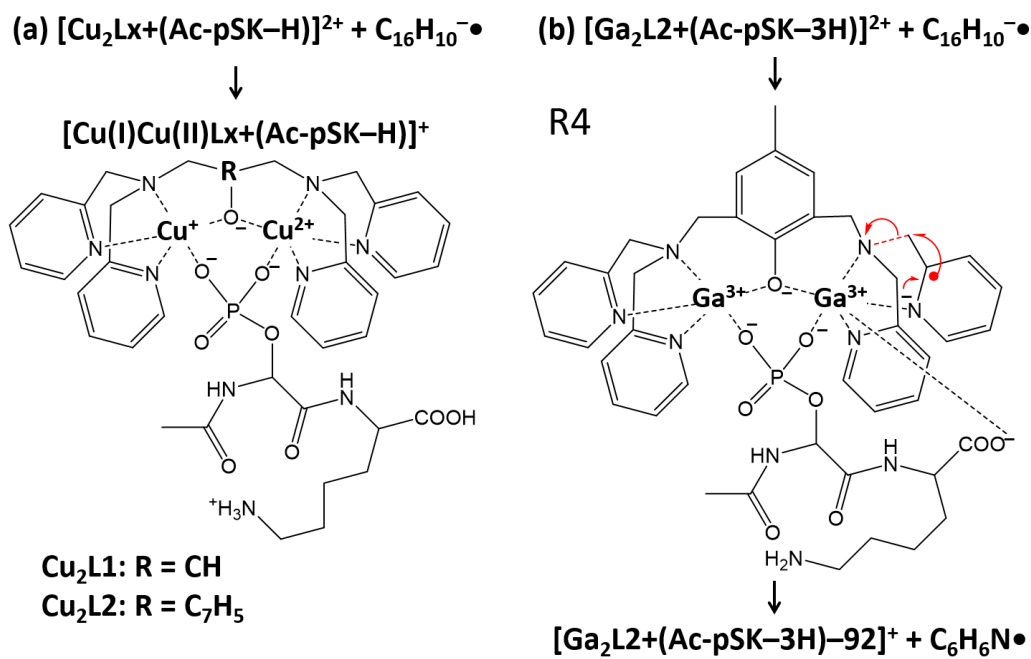

**Scheme 5. ETD mechanisms of (a) [Cu₂L1+(Ac-pSK–H)+H]²⁺ and (b) [Ga₂L2+(Ac-pSK–3H)]²⁺.**
**The detailed information including transition state geometry for the fragmentation of R4 were**
**shown in Supporting Information, Scheme S3.**

Figure 5d shows ETD-MS/MS mass spectrum of $[Ga_2L2 + (Ac\text{-}pSK - 3H)]^{2+}$, which selectively produced its fragment ion by $C_6H_6N\bullet$ loss. DFT calculations indicate that an electron selectively associates with the pyridine ring in the ligand and the subsequent geometry relaxation generates zwitterionic radical R4 (Scheme 5b). Then, R4 undergoes fragmentation by $C_6H_6N\bullet$ loss. According to the MN15/6-31++G(2d,p)//MN15/LanL2DZ/6-31G(d) level of calculation, the transition state energy barrier for $C_6H_6N\bullet$ loss is 71 kJ/mol, and the bond cleavage results in the formation of the IM, which was 19 kJ/mol more stable than R4. Subsequently, the complete dissociation energy of $[Ga_2L2 + (Ac\text{-}pSK - 3H) - 92]^+$ and $C_6H_6N\bullet$ is 127 kJ/mol above that of R4. The geometries of transition state barrier (TS) and intermediate (IM) for the corresponding fragmentation are shown in Supporting Information, Scheme S3. In summary, the use of $(Ga_2L2)^{5+}$ for ETD MS/MS preferentially provide 92-Da loss and did not induce peptide backbone fragmentation. Therefore, the $(Ga_2L2)^{5+}$-aided ETD are not suitable for phosphopeptide sequencing, although $(Ga_2L2)^{5+}$ strongly bind to phosphate group.

### The Utility of $(Zn_2L2)^{3+}$ for Phosphopeptide Sequencing by ETD-MS/MS

As described in the previous section, the use of $(Zn_2L2)^{3+}$ potentially facilitates phosphopeptide sequencing by ETD-MS/MS. Next, we used, T19p, and T43p as tryptic phosphopeptide models. These phosphopeptides have been used as models in previous studies, but the precise location of the phosphorylation site could not be determined because of the low sequence coverage when doubly protonated molecules were used as the precursors for ETD-MS/MS.[16] Figure 7 shows the $(Zn_2L2)^{3+}$-aided ETD-MS/MS mass spectra when triply charged $(Zn_2L2)^{3+}$-phosphopeptide complexes were used as the precursors. $(Zn_2L2)^{3+}$-aided ETD-MS/MS provided almost full sequence coverage, including information concerning the phosphorylation site. For comparison with the previously reported $(Zn_2L1)^{3+}$-aided ETD-MS/MS method, the change from $(Zn_2L2)^{3+}$ to $(Zn_2L1)^{3+}$ significantly suppressed the yield of the fragment arising from the 92-Da loss. In addition, the efficiency of N–Cα bond cleavage was improved, indicating that $(Zn_2L2)^{3+}$-aided ETD-MS/MS is a useful method for the sequencing of phosphopeptides.

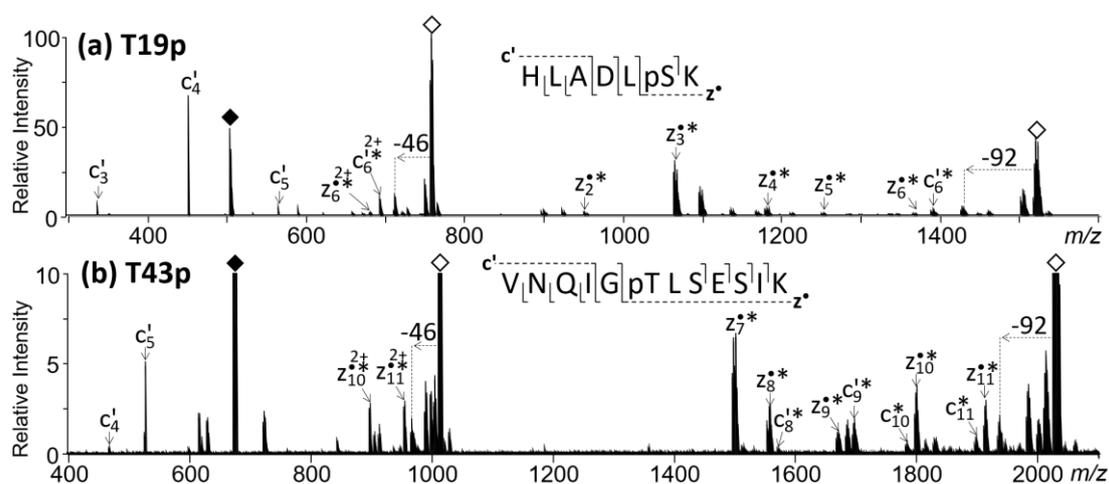

**Figure 7.** ETD-MS/MS mass spectra of triply charged $(Zn_2L2)^{3+}$-phosphopeptide complexes. The $c'$ and $z\bullet$ fragments annotated with asterisks ($c'*$ and $z\bullet*$) correspond to the $(Zn_2L2)^{3+}$ adduct on the $c'$ and $z\bullet$ fragments. Black and white squares indicate the precursor ions and charge-reduced products, respectively.

## CONCLUSION

The dinuclear metal complexes $(Zn_2L1)^{3+}$, $(Zn_2L2)^{3+}$, $(Cu_2L1)^{3+}$, $(Cu_2L2)^{3+}$, and $(Ga_2L2)^{5+}$ were investigated for use as additives for the analysis of phosphopeptides by ESI-based ETD-MS/MS. Although all dinuclear metal complexes selectively bound to phosphate compounds, their affinity is dependent on the type of metal present in the complex. The order of binding affinity for pS is $(Ga_2L2)^{5+}$ > $(Zn_2L1)^{3+}$ ≥ $(Zn_2L2)^{3+}$ > $(Cu_2L1)^{3+}$ > $(Cu_2L2)^{3+}$.

The type of metal ion in the complex also strongly influenced the ETD fragmentation. Regarding the ETD of the complexes with copper, $Cu^{2+}$ acted as an electron trap, and peptide fragmentation was largely suppressed. In contrast, $(Ga_2L2)^{5+}$ strongly bound to pS and phosphopeptides, whereas the ETD of the $(Ga_2L2)^{5+}$-phosphopeptide complex mainly led to $C_6H_6N\bullet$ loss. Therefore, the $(Cu_2L1)^{3+}$, $(Cu_2L2)^{3+}$, and $(Ga_2L2)^{5+}$-aided ETD-MS/MS could not be used for phosphopeptide sequencing.

The zinc dinuclear complexes, $(Zn_2L1)^{3+}$ and $(Zn_2L2)^{3+}$, efficiently bound to the phosphorylated compounds. In particular, the use of $(Zn_2L2)^{3+}$ facilitated N–Cα bond cleavage by ETD-MS/MS compared with the $(Zn_2L1)^{3+}$-aided method. The yield of the fragments arising from N–Cα bond cleavage reflects the number of aminoketyl radical intermediates, as predicted by TD-DFT calculations. The computational results indicate that the $(Zn_2L2)^{3+}$-phosphopeptide complex can efficiently produce aminoketyl radical intermediates by ETD, facilitating phosphopeptide sequencing. To apply $(Zn_2L2)^{3+}$-aided ETD for large-scale phosphopeptide analysis, the method should be combined with liquid chromatography mass spectrometry (LC-MS) system. Although the $(Zn_2L2)^{3+}$-phosphopeptide complex would not be stable during chromatographic separation process, the post-column addition of $(Zn_2L2)^{3+}$ can be produced the complex. The further investigation for $(Zn_2L2)^{3+}$-aided ETD combined with LCMS would promote wider application of this method.

## ACKNOWLEDGMENTS

DA acknowledge Issey Osaka, for allowing to use FT-ICR MS instrument. This work was supported by JSPS KAKENHI grant number 17K14508. The computations of molecular structures in this work were supported by Research Center for Computational Science, Okazaki and Center for Computational Sciences. The experiments were partly supported by Nanotechnology Platform Program of the Ministry of Education, Culture, Sports, Science and Technology (MEXT), Japan, and the Plateforme de Biophysico-Chimie Structurale of the IECB, France.